\documentclass[namedreferences]{SolarPhysics}
\usepackage[optionalrh]{spr-sola-addons} 
\usepackage{graphicx}
\usepackage{color}           
\usepackage{url}             

\newcommand{\aap}{    {\it Astron. Astrophys.}}
\newcommand{\aaps}{   {\it Astron. Astrophys. Suppl.}}

\newcommand{\apj}{    {\it Astrophys. J.}}
\newcommand{\apjl}{   {\it Astrophys. J. Lett.}}

\newcommand{\grl}{    {\it Geophys. Res. Lett.}}

\newcommand{\jgr}{    {\it J. Geophys. Res.}}

\newcommand{\solphys}{{\it Solar Phys.}}

\newcommand{\ssr}{    {\it Space Sci. Rev.}}

\begin{document}

\begin{article}

\begin{opening}

\title{Coronal Shock Waves, EUV Waves, and Their Relation to CMEs.
III. Shock-Associated CME/EUV Wave in an Event with a
Two-Component EUV Transient}

\author{V.V.~\surname{Grechnev}$^{1}$\sep
        A.N.~\surname{Afanasyev}$^{1}$\sep
        A.M.~\surname{Uralov}$^{1}$\sep
        I.M.~\surname{Chertok}$^{2}$\sep
        M.V.~\surname{Eselevich}$^{1}$\sep
        V.G.~\surname{Eselevich}$^{1}$\sep
        G.V.~\surname{Rudenko}$^{1}$\sep
        Y.~\surname{Kubo}$^{3}$
        }

\runningauthor{Grechnev et al.}
 \runningtitle{Shock-associated CME/EUV wave}

\institute{${}^{1}$Institute of Solar-Terrestrial Physics  SB RAS,
            Lermontov St.\ 126A, Irkutsk 664033, Russia email: \url{grechnev@iszf.irk.ru} \\
    $^{2}$  Pushkov Institute of Terrestrial Magnetism,
            Ionosphere and Radio Wave Propagation (IZMIRAN), Troitsk, Moscow
            Region, 142190 Russia email: \url{ichertok@izmiran.ru}\\
    $^{3}$ National Institute of Information and Communications
            Technology, Tokyo, Japan email: \url{kubo@nict.go.jp}
}

\begin{abstract}
On 17 January 2010, STEREO-B observed in extreme ultraviolet (EUV)
and white light a large-scale dome-shaped expanding coronal
transient with perfectly connected off-limb and on-disk
signatures. Veronig et al. (2010, ApJL 716, 57) concluded that the
dome was formed by a weak shock wave. We have revealed two EUV
components, one of which corresponded to this transient. All of
its properties found from EUV, white light, and a metric type II
burst match expectations for a freely expanding coronal shock wave
including correspondence to the fast-mode speed distribution,
while the transient sweeping over the solar surface had a speed
typical of EUV waves. The shock wave was presumably excited by an
abrupt filament eruption. Both a weak shock approximation and a
power-law fit match kinematics of the transient near the Sun.
Moreover, the power-law fit matches expansion of the CME leading
edge up to 24 solar radii. The second, quasi-stationary EUV
component near the dimming was presumably associated with a
stretched CME structure; no indications of opening magnetic fields
have been detected far from the eruption region.
\end{abstract}


\end{opening}


\section{Introduction}

Large-scale wave-like transients called EUV waves or ``EIT waves''
are observed in extreme ultraviolet (EUV) and soft X-rays in
association with coronal mass ejections (CMEs) and flares
(\citeauthor{Thompson1998}, \citeyear{Thompson1998,Thompson1999}).
Efforts of researchers to understand the nature of EUV waves meet
difficulties. The main observational material acquired with
SOHO/EIT suffers from insufficient temporal coverage. Bright flare
emission disfavors detection of faint EUV waves. Properties of
their propagation, association with flares and metric type II
bursts, \textit{etc.} appear to be diverse and contradictory
(\textit{e.g.}, \opencite{Biesecker2002}; \opencite{Klassen2000}).
Deficiency of observations stimulated development of conflicting
concepts based mainly on \textit{i})~MHD fast-mode disturbances
\cite{Thompson1999,Warmuth2001,KhanAurass2002,Long2008,Gopalswamy2009}
or \textit{ii})~plasma compression in bases of stretching loops
\cite{Delannee1999,Chen2002,Chen2005,Attrill2007} both caused by a
CME eruption (see also \opencite{Zhukov2004}; reviews of
\opencite{WillsDavey2009}; \opencite{Gallagher2010}). The former
set of hypotheses (\textit{i}) implies CME-related opening or
reconnection of magnetic fields in the vicinity of an eruption
site; the latter one (\textit{ii}) assumes it to be global to
describe both standing `EUV waves' and those propagating over
large distances.

The launch in 2006 of the twin-spacecraft
\textit{Solar-Terrestrial Relations Observatory} (STEREO;
\opencite{Kaiser2008}) carrying the Sun Earth Connection Coronal
and Heliospheric Investigation instrument suites (SECCHI;
\opencite{Howard2008}) significantly enhanced opportunities to
study EUV waves, including their temporal coverage, with the
Extreme Ultraviolet Imagers (EUVI). However, this has not lead to
consensus about their nature. Several studies argued the
shock-wave nature of observed EUV waves (\textit{e.g.},
\opencite{KienreichTemmerVeronig2009};
\opencite{PatsourakosVourlidas2009}; \opencite{Patsourakos2009}).
Conversely, \inlinecite{Zhukov2009} presented an EUV wave
incompatible with the fast-mode wave interpretation. Disappointing
was the study of the 19 May 2007 event, which was considered both
in favor of the shock-wave hypothesis
\cite{Long2008,VeronigTemmerVrsnak2008,Gopalswamy2009} and against
it \cite{Attrill2010,YangChen2010} [our analysis in Paper~I
\cite{Grechnev2011_I} supports the shock-wave interpretation]. A
recent analysis of an EUV wave observed in still more detail with
\textit{Solar Dynamics Observatory} provides more questions than
answers \cite{Liu2010}.

Diversity of EUV waves implies their probable relation to
different phenomena \cite{Zhukov2004,Grechnev2008,Cohen2009}. Our
companion Papers I\,--\,III consider EUV waves presumably
associated with coronal shock waves. Paper~I shows how to
reconcile shock-related EUV waves, type II bursts, and
corresponding CMEs. We propose a simple quantitative description
for all of these phenomena based on an approach of a self-similar
shock wave. The large length of such a wave is comparable with the
curvature radius of the wave front. Its deceleration is determined
by the increasing mass inside the volume limited by the shock
front. The self-similar approach describes propagation of strong
shock waves. Our experience has revealed a convenient way to fit
the kinematics of real coronal waves with direction-dependent
power-law (PL) functions (abbreviated henceforth `shock-PL fit').
\citeauthor{Afanasyev2011} (\citeyear{Afanasyev2011}; Paper~II)
have considered the opposite limit of a weak shock calculated
analytically in terms of the WKB
(Wentzel\,--\,Kramers\,--\,Brillouin) approach taking account of
nonlinear effects.

An eruptive event on 17 January 2010 produced a CME and wave,
whose expanding three-dimensional dome with its lower skirt
sweeping over the solar surface was observed in unprecedented
detail by EUVI and coronagraphs on the STEREO-B spacecraft. EIT
\cite{Delab1995} and LASCO \cite{Brueckner1995} instruments on
SOHO also observed this transient. A weak type II burst was
recorded by HiRAS (NICT, Japan) and Learmonth (US Air Force RSTN)
spectrographs. \inlinecite{Veronig2010} analyzed this backside
event and concluded that the coronal transient observed both in
EUV and white light was a dome of a `weakly shocked fast-mode MHD
wave'. The authors found that the lateral expansion of the wave
dome far from the eruption site was free, while, in their opinion,
its upward expansion was driven by the CME all the time.

We analyzed this event independently and also inferred the
shock-wave nature of this coronal transient. However, the scopes
and some conclusions of our and \inlinecite{Veronig2010} studies
do not coincide. Unlike the authors, we find deceleration of both
near-surface and off-limb traces of the wave. Our analysis shows
that the shock wave was most likely excited by the
impulsive-piston mechanism and freely propagated omnidirectionally
afterwards as considered in Paper~I. We study the shock wave
propagation both on-disk and off-limb in comparison with the
fast-mode speed ($V_\mathrm{fast}$) distribution and the drift
rate of the type II burst. We explain the differences between our
results and those of \inlinecite{Veronig2010}, and compare the
results, which the self-similar shock approximation and modeling
of a weak shock provide being applied to this event, including the
wave propagation at larger distances from the Sun. We have
revealed one more EUV transient, which adjoined the dimming and
was quasi-stationary. This fact confirms that different kinds of
``EIT waves'' do exist. We analyze the observations in
Section~\ref{S-observations}, compare the revealed properties of
the wave with modeling results in Section~\ref{S-discussion}, and
summarize the outcome in Section~\ref{S-summary}.

\section{Analysis of Observations}
 \label{S-observations}

The two STEREO spacecraft were located $69.6^{\circ}$ behind the
Earth and $64.3^{\circ}$ ahead of it. The eruption site shown by
the flare was seen from STEREO-B at S25~E59 (heliolatitude
$B0=3.74^{\circ}$) and located for observers on the Earth $\approx
37^{\circ}$ behind the east limb with a projected position onto
the visible solar surface of about S32~E55 ($B0=-4.75^{\circ}$).
The projected positions onto the visible solar surface observed
from the Earth (and SOHO) and STEREO-B were close to each other.
The radial CME extent and velocity were smaller by a factor of
1.13 for observers on the Earth (SOHO) with respect to
observations on STEREO-B. The STEREO-A/COR1 coronagraph registered
a wide transient around a position angle of $\mathrm {PA} \approx
225^{\circ}$ (we do not consider STEREO-A or EIT data).

\subsection{Eruption and a Probable Shock Wave}
 \label{S-profiles}

\begin{figure} 
  \centerline{\includegraphics[width=\textwidth]
   {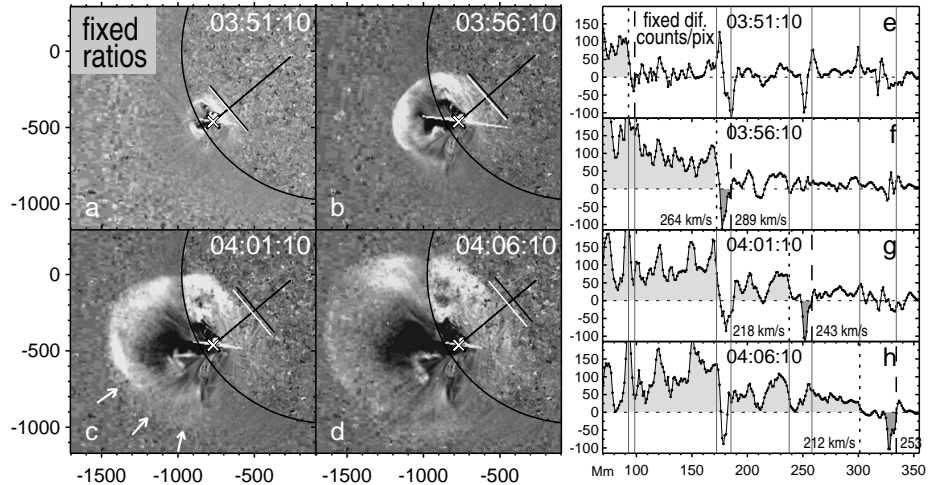}
  }
  \caption{a\,--\,d)~EUV wave and eruption in EUVI 195~\AA\ images.
White arrows in panel (c) indicate the fastest part of the front.
The slanted cross marks the eruption site. The black line going
from the eruption site northwest shows the direction where the
spatial profiles were computed. The white and black bars across
this line mark the presumable fronts suggested by the spatial
profiles. In all EUVI images hereafter, the axes show arc seconds
from the solar disk center as if viewed from the Earth.
e\,--\,h)~The spatial profiles of the EUV brightness measured in
the direction shown in the left panels. The vertical lines denote
presumable fronts. The shading indicates the EUV wave brightening
behind the front and a possible negative precursor ahead.
    }
  \label{F-euvi_prof}
  \end{figure}

Figure~\ref{F-euvi_prof}a\,--\,d and the movie euvi\_195.mpg in
the electronic version of our paper show the onset of the event
observed in EUVI 195~\AA\ fixed-base ratio images. A dome-like EUV
wave expanded above the limb and propagated along the solar
surface. The boundary of the surface front passed into the
off-limb dome suggesting their common nature. The front was
followed by extended brightenings indicating a large length of the
disturbance. Eruption and untwisting of a magnetic structure
(probably, a filament) is seen inside the EUV wave dome. The
motion of the eruption was three-dimensional. This fact, fading
out of the eruption, and difficulties to distinguish it from the
wave front make measurements of its kinematics unreliable.
Nevertheless, it is possible to see in
Figure~\ref{F-euvi_prof}a\,--\,d and in the movie that the
eruption changed shape like a straightening mainspring. Its
foremost edge was close to the wave front at 03:56. Thus, just an
abrupt eruption of the rope structure could have played a role of
an impulsive piston, which excited the wave, as was the case in
events considered in Paper~I.

The wave front is sharpest in Figure~\ref{F-euvi_prof}b\,--\,d and
Figure~\ref{F-front_v_fast}b just to the left from the eruption
site (slanted cross) in the plane of the sky, while the fastest
faint part of the front indicated by the white arrows in
Figure~\ref{F-euvi_prof}c is closer to the radial direction. The
sharper appearance of the front in the leftwards direction could
be due to overlap with expanding loops and a stronger shock in
this direction. The latter effect is consistent with the
predominant upwards motion of the eruption. The faintness of the
front in the South Pole's environment is due to the closeness of
the polar coronal hole, where the Alfv{\'e}n velocity
$V_{\mathrm{A}}$ is much higher (Figure~\ref{F-front_v_fast}h).
One more outcome is free propagation of the wave in the radial
direction, where the front moved ahead of possible loops
(Figure~\ref{F-euvi_prof}c).

Figure~\ref{F-euvi_prof}e\,--\,h shows plane-of-sky spatial
profiles computed from the four fixed-difference 195~\AA\ images
within sectors of $1^{\circ}$ along the directions indicated with
the black lines. The profiles show a relief constituted by
variations of compact features. The chosen direction crosses
features, which seem to have responded to the pass of the wave
front. The EUV wave brightening appears in the profiles as an
enhancement (light shading) to the left from the front (dotted).
All the profiles show compact darkenings (dashed, darker shading)
preceding the brightenings. The darkening and brightening regions
in Figure~\ref{F-euvi_prof}e seem to be imperfectly resolved,
which reduces the depth of the narrow darkening. The dashed and
dotted lines in Figure~\ref{F-euvi_prof}e\,--\,h correspond to the
black and white bars in Figure~\ref{F-euvi_prof}a\,--\,d.
Comparison of all the panels e\,--\,h with each other reveals
slightly variable compact features at the four fixed positions,
where the front presumably showed up. Hits by a shock front
probably disturbed the features, producing the sharp effect
suggested by the profiles, but not a gradual elevation. The EUVI
pixel size (small circles show the samples) was $ \approx 1190$
km; with exposure times of 16~s, a step-like front moving with
plane-of-sky speeds shown in Figure~\ref{F-euvi_prof}f\,--\,h must
be caught in 3\,--\,4 pixels. Thus, just such a response to a
shock front is expected.

Comparison of the profiles with the quiet Sun's level of about 290
counts/pixel shows that if this marginal effect was real, then its
value could only be due to disturbance of low structures. The
fact that the probable response of different solar features
matched arrival of the wave front at different times indicates
that the observed effect deserves attention to be checked in other
events.

After the pass of the wave front, small features like coronal
bright points got disturbed, but did not disappear (see also the
euvi\_195.mpg movie). This implies that closed magnetic fields in
these configurations did not open. No irreversible changes are
seen. No signatures of magnetic reconnection are detectable.

The plane-of-sky velocities of the presumable fronts
systematically decreased, despite their propagation from the
near-the-limb eruption site towards the solar disk center, that
must increase the projected speed. The surface velocities
estimated along an appropriate great circle all exceeded 390
km~s$^{-1}$ initially and all were less than 290 km~s$^{-1}$
finally, which indicates deceleration of the wave.

\subsection{Global EUV Wave Fronts}
 \label{S-global_fronts}

\begin{figure} 
  \centerline{\includegraphics[width=\textwidth]
   {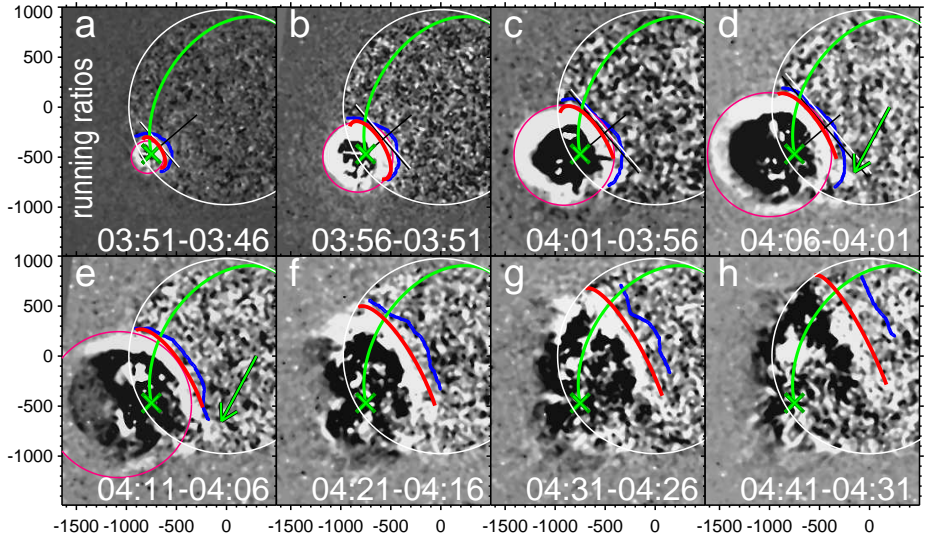}
  }
  \caption{Global wave fronts in EUVI 195~\AA\ images. The red arcs
outline the leading edges of the global wave fronts detectable in
the figure. The blue lines correspond to the fronts of Veronig
\textit{et al.} (2010). The pink ovals outline the off-limb wave
dome. The arrows indicate a bend of the fronts into the South
Pole's coronal hole. Distances along the solar surface were
measured from the eruption cite (the slanted cross) along the
green great circle. The white and black bars in panels a\,--\,d
mark the presumable fronts from Figure~\ref{F-euvi_prof}.
    }
  \label{F-global_euv_wave_fronts}
  \end{figure}

We divided the problem of identifying the wave fronts into two
tasks: 1)~identification of global fronts and 2)~analysis of
smaller-scale properties of the EUV wave propagation (next
Section). We reveal global wave fronts from ratios of
running-difference images to preceding ones. To detect weak
portions of the fronts, the images were rebinned to $512 \times
512$ pixels and deeply filtered using smoothing with a boxcar, a
median smoothing, and displayed by means of the \textit{sigrange}
SolarSoftware routine. The result is shown in
Figure~\ref{F-global_euv_wave_fronts} (eight of 12 images used
with a total coverage of 55 min). We separately outlined the
on-disk and off-limb parts of the fronts with red and pink ovals,
trying to catch their outermost envelopes over a maximal spatial
extent. The distances were measured along the green great circle.
The technique used by \inlinecite{Veronig2010} was more sensitive.
They analyzed spatial profiles computed within some sectors and
searched for their foremost edges close to the visually identified
fronts. The blue contours approximately reproduce the fronts,
which the authors identified.

Figure~\ref{F-global_euv_wave_fronts} shows the following facts.
1)~The fronts identified by \citeauthor{Veronig2010} all lead our
fronts with increasing separation. 2)~The southern part of the
front indicated by the arrows in
Figure~\ref{F-global_euv_wave_fronts}d,\,e moved considerably
faster in the environment of the polar coronal hole, while the
fronts themselves were difficult to detect there. 3)~The wave dome
expanded non-radially: with the southeastern position of the
eruption site, expansion of the dome was pointed almost exactly to
the left in the plane of the sky. Also, the projection of the
off-limb dome center onto the solar surface increasingly shifted
northeast, so that the fronts in later images were not parallel to
the earlier ones (\textit{cf., e.g.,}
Figure~\ref{F-global_euv_wave_fronts}d and
\ref{F-global_euv_wave_fronts}h).

The lag of the red fronts behind the blue ones is initially small
and nearly constant, and then increases. Since the speed of the
blue fronts was constant \cite{Veronig2010}, this behavior implies
deceleration of the red fronts. Indeed, our distance-time plots in
Figures \ref{F-spectrum_fit}b,
\ref{F-calculated_dome_and_kinematics}b (red symbols) show
deceleration. The plots are well fitted with PL functions expected
for a shock wave (see Paper~I):
\begin{eqnarray}
x(t) = x_1[(t-t_0)/(t-t_1)]^{\alpha},
 \label{E-pl_fit}
\end{eqnarray}
where $t$ and $x$ are current time and distance, $t_0$ = 03:47:48
is the wave start time (estimated in sequential attempts to fit
the EUV wave and the type II burst), $t_1$ and $x_1$ correspond to
one of the measured fronts, and the PL exponent $\alpha =
2/(5-\delta)$ with $\delta$ being a density falloff index in this
formal approximation. We fitted the kinematics of the wave front
with an exponent of $\alpha \approx 0.75$ ($\delta \approx 2.35$)
for the surface propagation and $\alpha \approx 0.91$ ($\delta
\approx 2.80$) for the off-limb expansion.

The measurements of the velocities along the great circle have
largest uncertainties at earliest stages of the wave expansion,
and for the initial interval of 15 min we also used 171~\AA\
images observed with a higher imaging rate. The velocity
corresponding to a power-law distance-time plot has a singularity
in the origin $t_0$ and is not limited from below by
$V_\mathrm{fast}$ at large distances. Hence, the $\delta$
parameter is expected to be slightly different for long and short
time intervals beginning with $t_0$ (actually 55 min for 195~\AA\
images and 15 min for 171~\AA\ ones).

\begin{figure} 
  \centerline{\includegraphics[width=\textwidth]
   {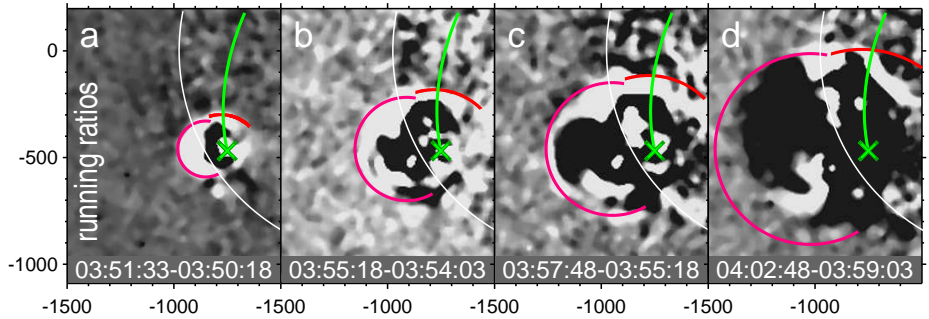}
  }
  \caption{Global wave fronts detectable in EUVI 171~\AA\
images and their outline. The red arcs outline the leading edges
of the on-disk wave fronts detectable in the figure. The pink
ovals outline the off-limb wave dome. The green arc denotes a
great circle along which our measurements were made. The slanted
green cross marks the eruption site.
    }
  \label{F-euvi_171_fronts}
  \end{figure}

Figure~\ref{F-euvi_171_fronts} presents four of 12 EUVI 171~\AA\
images which we used. Since the shock-PL fit applies, we used its
parameters found from the 195~\AA\ data as an initial estimate and
endeavored to outline each of the on-disk and off-limb wave
portions with ovals calculated from the shock-PL fit according to
the observation times at 171~\AA. The $\delta$ parameter was
adjusted to match the fronts. If some parts of the fronts were not
detectable, we used their other possible signatures. An extreme
example is shown in Figure~\ref{F-euvi_171_fronts}d. Here the
reference regions for the off-limb oval were the upper (in the
plane of the sky) brightening just above the limb and three faint
lowermost compact regions. The on-disk oval was referred to the
bright feature crossing the limb and a small portion of the front
next to the former feature. The results of the measurements shown
in Figures~\ref{F-spectrum_fit}b and
\ref{F-calculated_dome_and_kinematics}b with blue triangles are
fitted with $\delta = 2.74$ for the off-limb dome and $\delta =
2.1$ for the surface propagation (blue curve). The difference
between $\delta$ found from the 195~\AA\ and 171~\AA\ images is
detectable in the velocity-time plot in
Figure~\ref{F-calculated_dome_and_kinematics}c.

The measurements based on outline of entire wave fronts reveal
some indistinct effects such as the motion of the wave center and
provide tighter uncertainties than, \textit{e.g.}, measurements of
a fastest front portion do. Preconditioning with an expected
analytic function still tightens the uncertainties. Estimating
them is not a simple task. One way is to find the extreme
envelopes enclosing possible options of the outline, but it is
time consuming. We alternatively assumed extreme uncertainties of
the front identification of 1.5 minor ticks (110 Mm) in the latest
195~\AA\ images and twice lesser ones at 03:56. The uncertainty in
$t_0$ estimated from the type II burst was assumed to be $\pm
30$~s. The resulting gray scatter bands in
Figure~\ref{F-calculated_dome_and_kinematics}b,\,c for the surface
distance and velocity plots vs. time correspond to $\delta = 2.35
\pm 0.05$. The assumed uncertainties of the front identification
appear to be well overestimated; realistic bands should be
considerably narrower.

\subsection{EUV Wave Components and Fast-Mode Speed Distribution}
 \label{S-euv_wave_fast_mode_speed}

\begin{figure} 
  \centerline{\includegraphics[width=\textwidth]
   {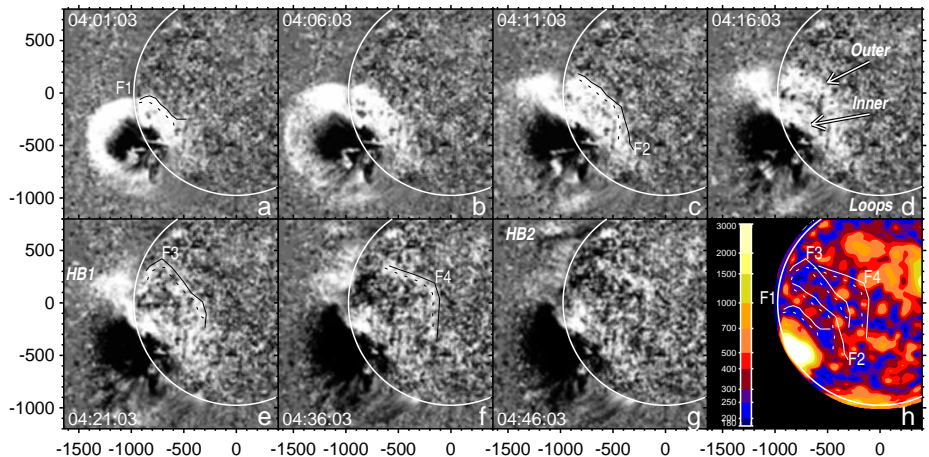}
  }
  \caption{The EUV wave in fixed-base ratio EUVI 195~\AA\ images (a\,--\,g) and
$V_{\mathrm{fast}}$ distribution at 30 Mm (h). The white circles
outline the solar limb. The stationary inner component and the
propagating outer one are denoted in panel (d). Some portions of
the wave fronts are outlined with black contours F1\,--\,F4 at
their foremost (solid) and brightest (dotted) parts. The scale bar
in panel h quantifies the $V_{\mathrm{fast}}$ levels in
km~s$^{-1}$.
 }
  \label{F-front_v_fast}
  \end{figure}

Running difference images are best suited to emphasize outermost
fronts, but inner quasi-stationary features do not show up in such
images. The EUV wave in this event is well visible in fixed-base
ratio EUVI 195~\AA\ images in Figure~\ref{F-front_v_fast}a\,--\,g
allowing us to see what happened behind the expanding front. The
whole large-scale brightening consisting of small patches was wide
and complex. The outer propagating front included another, inner
EUV transient. After an initial evolution, its on-disk part
adjoining the dimming became stationary. Its brightness initially
was comparable with the outer front and exceeded it later on. The
inner component appears to have consisted of two parts
distinguishable in Figure~\ref{F-front_v_fast}f,g and in the
movie, with the northern part slowly moving northeast.

An off-limb brightening (HB1) visible up to 290~Mm above the inner
component slowly expanded northwards along the limb for about half
an hour following the outer front. This brightening could be due
to a portion of the outer front sweeping over the backside part of
the solar surface.

A high-altitude brightening HB2 (Figure~\ref{F-front_v_fast}f,\,g)
at about 230 Mm appeared when the wave front reached it suggesting
that a lower dense part of a coronal streamer highlighted the
front. Long loops connecting the active region with a southwestern
area (`Loops' in Figure~\ref{F-front_v_fast}d) also highlighted
the wave. Most of the loops outside of the active region did not
show any stretch, although the wave front passed through these
loops. These facts support a wave nature of the outer EUV wave.
There are no manifestations of magnetic field opening aside of the
eruption region, where, however, the outer EUV wave was visible.
The limited magnetic field opening is also confirmed by the
confined coronal dimming region in Figure~\ref{F-front_v_fast}
that was first stated by \inlinecite{Veronig2010}. Just the
stationary inner brightening appears to be related to a stretched
CME structure.

The conclusion of \inlinecite{YangChen2010} that `...EIT wave
propagates more slowly in the regions of stronger magnetic field'
inspired us to compare near-surface EUV wave manifestations with
the $V_{\mathrm{fast}}$ distribution ($V_{\mathrm{fast}}^2 \approx
V_{\mathrm{A}}^2 + C_{\mathrm{s}}^2$ with the sound speed
$C_{\mathrm{s}}$ is assumed to be 180~km~s$^{-1}$ everywhere). We
calculate $V_{\mathrm{fast}}$ from the \textit{magnitude}
$|\mathbf{B}|$ of the magnetic field, which determines the
Alfv{\'e}n speed rather than any magnetic component. The magnetic
field was extrapolated to 30 Mm from a SOLIS magnetogram observed
at 19:30 on 20 January using potential approximation
\cite{Rudenko2001}. A simplest way to obtain a $V_{\mathrm{fast}}$
distribution is to assume a constant temperature and to take
densities, \textit{e.g.}, from the Saito model. We attempt to get
a somewhat more realistic density distribution using a SOHO/EIT
195~\AA\ image observed on 20 January and an expression $\log n_e
= 8.34 + 0.509\log I_{195}$ obtained by \inlinecite{Brosius2002}
in a study of a particular region ($n_e$ is the electron density,
$I_{195}$ is the brightness in the 195~\AA\ EIT channel). This
expression cannot be universal, because the EUV brightness,
$I_{195} \propto n_e^2L$, depends on the depth $L$. However,
$V_{\mathrm{A}}$ depends on the depth weakly, $\propto L^{1/4}$,
and we restricted the density above quiet regions by limiting
plasma beta $\beta \leq 0.65$ (see, \textit{e.g.},
\opencite{WarmuthMann2005}). The resulting $V_{\mathrm{fast}}$
distribution is presented in Figure~\ref{F-front_v_fast}h (the
highest-speed values above the active region are limited by $\leq
3000$~km~s$^{-1}$ to reveal low-speed regions throughout the solar
disk). This distribution is not accurate for the following
reasons. Most likely, the high-speed area in the active region was
smaller on 17 January than the three-days later magnetogram shows.
The density could be underestimated there, thus somewhat
increasing $V_{\mathrm{A}}$. We cannot untangle the height
dependence of the density distribution from an EUV image of the
solar disk. These inaccuracies are not essential for our results.

The shock formation time can be estimated from this distribution.
A disturbance caused by an impulsive filament eruption steepens
into a shock presumably in a region of a sharp falloff of
$V_{\mathrm{fast}}$ due to jam of the wave profile. With a
half-width of the high-speed area above the active region (white
area in Figure~\ref{F-front_v_fast}h) of about 100~Mm, the shock
must form in the lateral direction within one minute [consistent
with the estimate of \inlinecite{Grechnev2008} for a different
event]. With the wave onset time $t_0 \approx$ 03:47:48, this
estimate is consistent with the fact that the type II burst was
observed as early as 03:51 indicating that the shock already
existed in the upwards direction at that time.

The on-disk EUV wave was distinct in an area between the active
region and a large northeastern high-speed domain.
$V_{\mathrm{fast}} = 290$~km~s$^{-1}$ dominated there. The surface
EUV wave speed was from 300\,--\,325~km~s$^{-1}$ at 04:01 to
240\,--\,270~km~s$^{-1}$ (see
Figure~\ref{F-calculated_dome_and_kinematics}c) at it latest
observation. Thus, the near-surface portion of the wave front was,
most likely, in the weak shock regime in regions of low
$V_{\mathrm{fast}}$ and propagated almost as a linear fast-mode
wave in regions of increased $V_{\mathrm{fast}}$.

Four distinct portions of the EUV wave fronts F1\,--\,F4 are
outlined both in EUVI images
(Figure~\ref{F-front_v_fast}a,\,c,\,e,\,f) and on the
$V_{\mathrm{fast}}$ distribution (Figure~\ref{F-front_v_fast}h).
The solid lines trace the foremost fronts; the dotted lines trace
their brightest parts. Comparison shows that the EUV wave was
brightest in regions of lowest $V_{\mathrm{fast}}$. The boundary
of the EUV wave corresponded to regions, where $V_{\mathrm{fast}}$
increased. Portions of the fronts located in regions of increased
$V_{\mathrm{fast}}$ stretched and lost brightness. Fronts
sharpened, brightened and suspended motion in regions of low
$V_{\mathrm{fast}}$. Some suspended front portions are detectable
in two or even more images. Practically the same front F4 persists
in Figure~\ref{F-front_v_fast}f,\,g. The euvi\_195.mpg movie shows
other examples. The southern branch of front F2 is detectable at
04:06\,--\,04:16. The northern bend of front F3 is visible at
04:21 and 04:26. That is, the small-scale $V_{\mathrm{fast}}$
distribution did not determine the overall kinematics of the wave
(see Figure~\ref{F-global_euv_wave_fronts}), but affected the
brightness and sharpness of the wave front. Indeed, the Mach
number $M = V_{\mathrm{shock}}/V_{\mathrm{fast}}$ increases in
regions of reduced $V_{\mathrm{fast}}$, $\Delta M \approx
-(M-1)\Delta V_{\mathrm{fast}}/V_{\mathrm{fast}}$, \textit{i.e.},
the plasma compression is stronger.

These facts agree with the perturbation profile evolution revealed
by \inlinecite{Veronig2010} in averaging over a spherical sector
of $60^{\circ}$: the profile initially increased in magnitude and
sharpened until 04:01, and thereafter evolved in the reverse
manner. Since the shock most likely appeared 10 min before 04:01,
the observed steepening was not due to the shock formation. The
sharpest and brightest front found by the authors at 04:01
corresponds to our front F1 (Figure~\ref{F-front_v_fast}a,~h),
which was mostly aligned with a deep valley in the
$V_{\mathrm{fast}}$ distribution. \inlinecite{Veronig2010}
estimated the Mach number averaged over the F1 front to be 1.15;
we estimate that locally it could reach 1.5. Later on, the shock
probably dampened, as the authors concluded. Besides, the depth,
homogeneity, and orientation of each subsequent valley relative to
the wave front and to the measurement direction were not as
favorable as in the first valley. Dispersion of the front over
increasing width of the authors' measurement sector also
increased.

Thus, kinematics of the shock wave was governed by the large-scale
$V_{\mathrm{fast}}$ distribution: the wave ran much faster in the
polar coronal hole and its environment. Conversely, when the wave
expanded enough to exceed compact structures, the effect of the
small-scale near-surface distribution of $V_{\mathrm{fast}}$ was
more pronounced in sharpness and brightness of the EUV wave than
in its local speed.

The two EUV wave components in this event remind us of two
disturbances in Fig.~7 from \inlinecite{Chen2005}. One disturbance
is a compressive effect due to opening magnetic fields during the
CME lift-off as initially proposed by \inlinecite{Delannee1999}.
This disturbance moves slowly and stops at a magnetic separatrix.
Such influence of the CME lift-off on magnetic fields is expected
to diminish at distances well exceeding the initial size of an
eruptive magnetic rope suggested by the post-eruptive arcade and
the major deep stationary dimmings nearby (if the CME does not
involve magnetic fields connected to remote active regions). The
second, a faster disturbance is a trail of a coronal shock wave
sweeping over the solar surface. With the qualitative and
quantitative properties of the outer propagating disturbance
revealed by \inlinecite{Veronig2010} and in this Section, it is
difficult to imagine an alternative to its interpretation as a
trail of a coronal shock wave. The presence of both predicted wave
and non-wave components of the EUV transient in this event offers
a promising opportunity to settle debates over the nature of EUV
waves.

\subsection{White-light Coronal Transient Observed with COR1 and LASCO/C3}
 \label{S-wl_transient}

\begin{figure} 
  \centerline{\includegraphics[width=\textwidth]
   {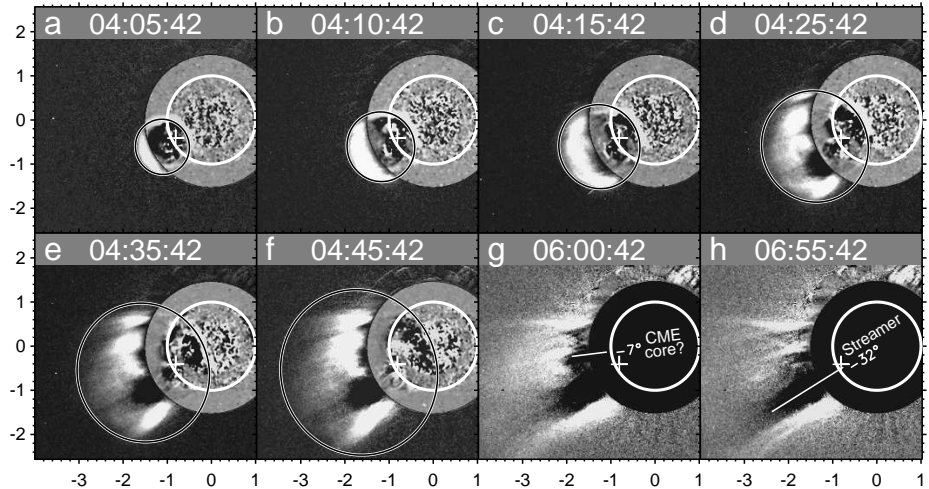}
  }
  \caption{Coronal transient in STEREO-B/COR1 fixed
difference images with inserted EUV wave fronts in
running-difference EUVI images (a\,--\,f) from
Figure~\ref{F-global_euv_wave_fronts}. Thick white circles denote
the solar limb. Black-on-white ovals outline the CME edge
according to the shock-PL fit. The cross marks the eruption site.
The axes show distances from the solar disk center in solar radii.
    }
  \label{F-cor1}
  \end{figure}

Figure~\ref{F-cor1} shows eight STEREO-B/COR1 images of a coronal
transient. An image observed at 03:50 was subtracted from all
others. The ovals outlining the edge of the transient correspond
to the shock-PL fit with the same $t_0$ = 03:47:48. To coordinate
the ovals with the non-radially expanding transient
(Section~\ref{S-global_fronts}), their centers are increasingly
shifted and different expansion factors are used in the radial
direction $\delta_{\mathrm{rad}} = 2.80$ and the transversal one
$\delta_{\mathrm{trans}} = 2.85$, \textit{i.e.}, the front tended
to become oblate. The ovals match the fronts in EUVI images and
cling to the outermost edges of the transient. The leading edge
decelerated as a freely propagating shock wave. The foremost part
looks like a plasma flow streaming along the fan of coronal rays.
CME structures are surmised well behind the leading edge. This
picture suggests the plasma flow successively involved into the
motion by a freely propagating shock front, whose speed was the
phase velocity of the involvement. A structure at an angle of
$-7^{\circ}$ ($\mathrm{PA} = 97^{\circ}$) in
Figure~\ref{F-cor1}g,\,h might be the CME core. The
density-depleted major streamer appears in difference images as a
wide radial darkening around $-32^{\circ}$ ($\mathrm{PA} =
122^{\circ}$).

  \begin{figure} 
  \centerline{\includegraphics[width=\textwidth]
   {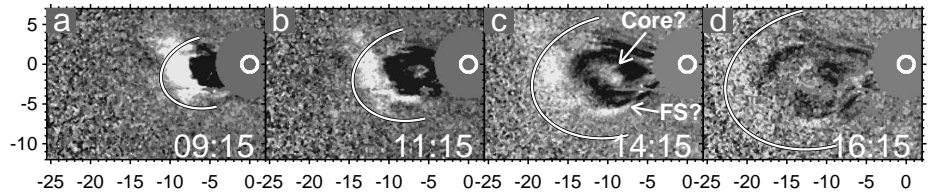}
  }
  \caption{The CME observed with LASCO/C3 (running differences). The thick white
circle denotes the solar limb. The white-on-black ovals outline
the outermost CME edge according to the shock-PL fit. The axes
show distances from the solar disk center in solar radii.
    }
  \label{F-lasco_c3}
  \end{figure}

We consider LASCO/C3 images only (C2 data became available later).
Figure~\ref{F-lasco_c3} presents four of 12 images, in which the
CME is detectable. Probable frontal structure and core are
indicated in Figure~\ref{F-lasco_c3}c. A ragged, diffuse
presumable plasma flow seems to be present ahead. We outlined the
CME with an oval corresponding to expansion of the shock wave. The
ovals in Figure~\ref{F-lasco_c3} calculated according to the
shock-PL fit with the same start time $t_0=$ 03:47:48 and $\delta
= 2.80$ embrace the CME but the fastest feature at $\mathrm{PA}
\approx 70^{\circ}$, most likely of a non-wave nature. Rather poor
observations and the low CME speed ($< 400$ km~s$^{-1}$) do not
allow us to find out if the change of its shape was due to effects
of the shock propagation or acceleration of the solar wind.
Nevertheless, the shock-wave kinematics does not contradict even
LASCO/C3 observations up to $23R_{\odot}$.

\subsection{Expansion of the Wave Dome and the Type II Burst Drift Rate}
 \label{S-type_ii_burst}

\begin{figure} 
  \centerline{\includegraphics[width=\textwidth]
   {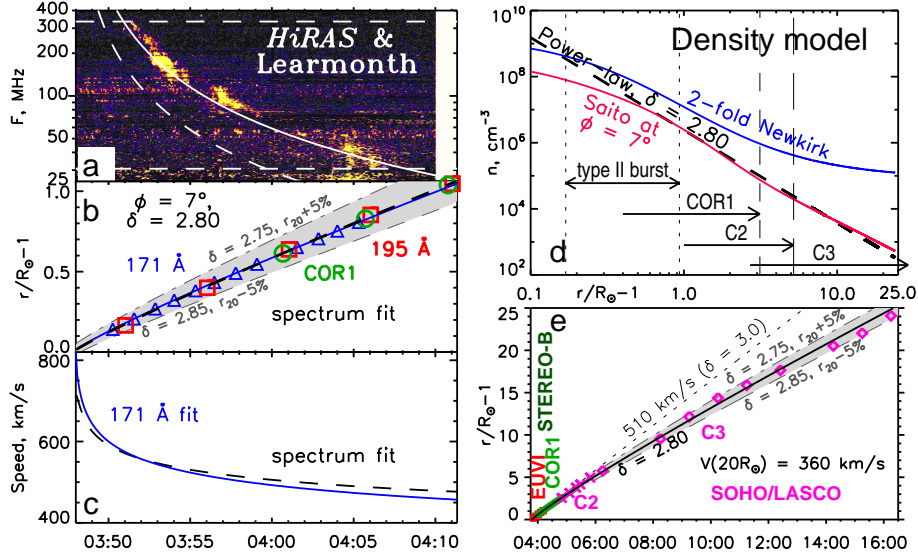}
  }
  \caption{Off-limb expansion of the EUV wave and type II burst.
a)~Composite dynamic spectrum. b)~Height-time measurements
(symbols) from EUVI at 195~\AA\ (red), 171~\AA\ (blue), and COR1
(green); shock-PL fit (blue line) and fit of the type II burst
converted into heights (dashed black). c)~Speed-time plots
calculated from the shock-PL fit of the 171~\AA\ data (blue) and
from the dynamic spectrum (dashed black). d)~Coronal density
models: PL model with $\delta = 2.8$ (dashed black) fitting the
dynamic spectrum and models of Newkirk (blue) and Saito for $\phi
= 7^{\circ}$ (red). e)~Overall height-time plot including the CME
Catalog data (pink) embraced by the gray band (also shown in panel
b). The dotted line is a linear fit of data in panel (b).
    }
  \label{F-spectrum_fit}
  \end{figure}

Figure~\ref{F-spectrum_fit}a shows a dynamic spectrum combined
from HiRAS and Learmonth records. The type II burst had a single
band most likely corresponding to the second harmonic, because the
fundamental emission must be strongly refracted due to the far
backside location of the eruption site. \inlinecite{Veronig2010}
came to the same conclusion. The drift of the burst is well
outlined with the PL model $n = 5.5 \times 10^8 (h/100\
\mathrm{Mm})^{-2.8}$ and the same wave start time $t_0$ =
03:47:48. The dashed outline corresponds to a presumable
fundamental emission.

We consider the direction $\vartheta = -7^{\circ}$ matching the
sharpest part of the EUV wave front non-radially expanding above
the limb and a probable CME core in Figure~\ref{F-cor1}g.
Figure~\ref{F-spectrum_fit}b shows measurements of the wave dome
from EUVI 171~\AA\ and 195~\AA\ images along with a shock-PL fit
of the 171~\AA\ data and the frequency drift converted into the
height-time plot. Figure~\ref{F-spectrum_fit}c shows speed-time
plots corresponding to the 171~\AA\ fit and the type II burst.
Figure~\ref{F-spectrum_fit}d presents our PL density model, the
\inlinecite{Newkirk1961} model, and the \inlinecite{Saito1970}
model at $ \phi \approx |\vartheta| = 7^{\circ}$. The PL model is
close to the 2-fold Newkirk model at the onset of the type II
burst and later approaches the Saito model. The difference with
the Newkirk model here, unlike the events addressed in Paper~I,
might be due to the non-radial wave expansion. The arrows show the
height ranges corresponding to the type II burst and fields of
view of coronagraphs.

Figure~\ref{F-spectrum_fit}e shows an overall height-time plot
including measurements from the SOHO LASCO CME Catalog
\cite[\url{http://cdaw.gsfc.nasa.gov/CME_list/}]{Yashiro2004} at
$\mathrm{PA} = 97^{\circ}$ up to $24R_{\odot}$. To coordinate the
measurements at different position angles from two different
vantage points of STEREO-B and SOHO, all the distances from the
CME Catalog are increased by 4.2\%. The shock-PL fit with $\delta
= 2.80$ corresponding to the Saito model matches all the data. The
boundaries of the gray band in Figure~\ref{F-spectrum_fit}b,\,e
covering all the measurements in the Catalog correspond to $x_1 =
20R_{\odot} \pm 5\%$, $\delta = 2.80 \pm 0.05$, and $t_0 =
$~03:47:48~$\pm 30$~s in expression (\ref{E-pl_fit}). This band
presents the scatter of measurements in the CME Catalog in terms
of uncertainties of the shock-PL fit.

Deceleration of the wave is not pronounced within $1.1R_{\odot}$
(Figure~\ref{F-spectrum_fit}b). This explains why
\inlinecite{Veronig2010} found a constant wave speed of
650~km~s$^{-1}$ in the radial direction ($\vartheta \approx
-32^{\circ}$). The linear-fit speed in the direction $\vartheta
\approx -7^{\circ}$ was 510~km~s$^{-1}$. Deceleration of the wave
becomes detectable from COR1 measurements up to $3R_{\odot}$ from
the eruption site. In fact, this determines the measurement
accuracy of $\delta$ ($\delta = 3.0$ for the constant speed). The
dotted line in Figure~\ref{F-spectrum_fit}e is a constant-speed
plot extrapolating the linear fit in Figure~\ref{F-spectrum_fit}b.

The fact that the shock-PL fit matches expansion of the slow
coronal transient up to $24R_{\odot}$, where its speed became
comparable with the solar wind speed, suggests that the leading
wave and the trailing piston were not completely independent.
Synergy between the piston and wave discussed in Paper~I (Section
4.3) is indeed expected to become pronounced at large distances
from the Sun.

\section{Discussion}
 \label{S-discussion}

The detailed STEREO/SECCHI observations of the EUV wave allow us
to compare the results of the shock-PL fit proposed in Paper~I
with those of the weak shock modeling described in Paper~II
(hereafter WS model). The EUV wave propagated mainly over quiet
Sun's regions without large-scale features except for the polar
coronal hole. Since the EUV wave most likely was a near-surface
trail of a large-scale coronal MHD wave, its kinematics should not
be significantly affected by small-scale inhomogeneities, as the
observations confirm (Section \ref{S-euv_wave_fast_mode_speed}).
We describe the global propagation of the EUV wave outside of the
active region assuming only radial variations of coronal plasma
parameters. The on-disk EUV wave decelerated from $\gsim 390$ to
$\lsim 290$ km~s$^{-1}$ (Section \ref{S-profiles});
\inlinecite{Veronig2010} found broadening of the wave profile.
These facts along with estimates of $V_{\mathrm{fast}}$ in the
lower corona above the quiet Sun indicate that the shock was weak
to moderate, so that the WS model appears to apply.

The model is not yet able to incorporate coronal magnetic fields
extrapolated from real magnetograms. We therefore model kinematics
of only an on-disk wave running over the quiet Sun. We use the
barometric density falloff of isothermal coronal plasma
$n(r)=4\times 10^8 \exp \left\{ 9.71\,(R_{\odot}/r-1) \right\} $
cm$^{-3}$ with coronal temperature $T=1.5$ MK ($C_{\mathrm{s}} =
180$~km~s$^{-1}$), and the radial magnetic field model
$B_r=1.35\,(R_{\odot}/r)^2$~G. $V_{\mathrm{A}} = 170$ km~s$^{-1}$
at 40 Mm and increases upwards. We assume that the wave originates
at an initial surface as large as 200 Mm, inside which the wave
source is located. Then an EUV wave front can be observed at
03:51. The model shock wave has an initial length of 80 Mm and an
amplitude of $1.5\,V_{\mathrm{fast}\,0}$ ($V_{\mathrm{fast}\,0}$
corresponds to the source height of 80 Mm). We search for EUV
signatures of the coronal wave at a height of 40 Mm.

\begin{figure} 
  \centerline{
                \includegraphics[width=0.5\textwidth,clip=]{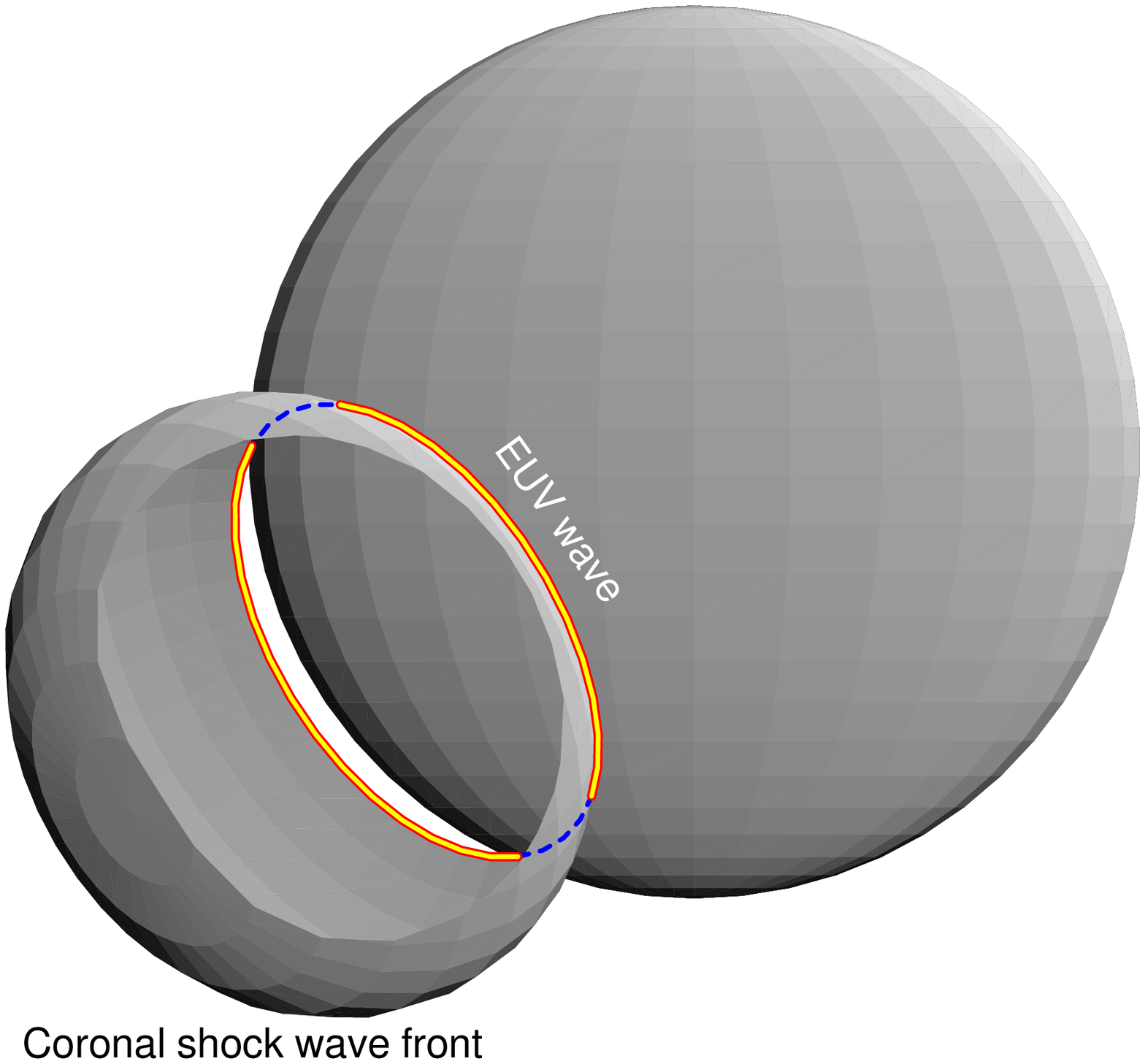}
                \includegraphics[width=0.5\textwidth,clip=]{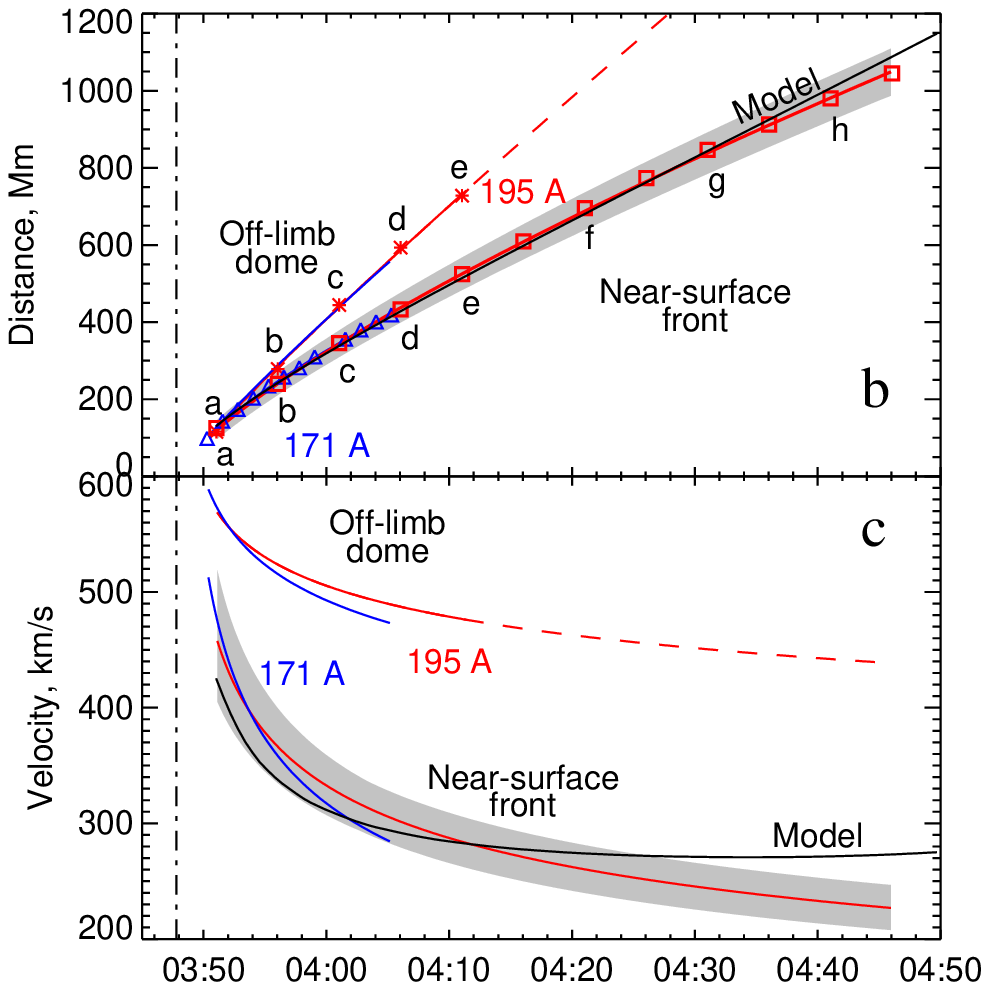}
}
    \vspace{-0.10\textwidth}
          \centerline{\large
      \hspace{0.4 \textwidth}  {a}
          \hfill}
   \vspace{0.075\textwidth}
   \caption{Measurements, fit, and modeling of shock front propagation.
a)~The modeled shock front. b)~Distance-time plots of the on-disk
wave (measured along the great circle in
Figure~\ref{F-global_euv_wave_fronts}) and the off-limb dome,
their shock-PL fit, and a modeled plot (195~\AA\ red, 171~\AA\
blue, model black; the labels a\,--\,h denote the corresponding
frames in Figure~\ref{F-global_euv_wave_fronts}). c)~The same for
the calculated velocities. The gray bands present extreme
uncertainties discussed in Section~\ref{S-global_fronts}.
    }
  \label{F-calculated_dome_and_kinematics}
  \end{figure}

Figure~\ref{F-calculated_dome_and_kinematics} shows some results
of the modeling and the measurements.
Figure~\ref{F-calculated_dome_and_kinematics}a presents the
computed 3D shock front. The color rim is the section of the wave
dome at 40 Mm. The on-disk EUV front could be partly covered by
the dome. Figure~\ref{F-calculated_dome_and_kinematics}b shows the
distance-time plots of the on-disk EUV front measured at 195~\AA\
(red squares) and 171~\AA\ (blue triangles). The red and blue
lines show the corresponding shock-PL fit; the black line presents
model results. Figure~\ref{F-calculated_dome_and_kinematics}c
shows the velocity-time plots obtained by differentiating of the
shock-PL curves and the modeled plot. The EUV wave appreciably
decelerates due to damping and then slightly accelerates because
of an increasing tilt of the front to the solar surface that is
discussed in Paper~II. The deviation of the speed supplied by the
shock-PL fit from the result of the WS modeling does not exceed
15\%.

We have also modeled propagation of a shock wave upwards. Active
regions determine a $V_{\mathrm{A}}$ distribution in their
vicinities. To simulate this effect, we have added a radial
magnetic dipole into our radial magnetic field model as
\inlinecite{WarmuthMann2005} did. A `horizontal' dipole seems to
conform to the active region on 17 January. Embedding such a
dipole into the model results in strongly anisotropic
$V_{\mathrm{fast}}$ distribution in the corona with a domain of
very low $V_{\mathrm{fast}} \approx C_{\mathrm{s}}$ near a null
point of the magnetic field and that of enhanced
$V_{\mathrm{fast}}$. This causes asymmetric wave front propagation
actually observed in this event. However, the domain of influence
of a `horizontal' dipole is too large, comparable with the solar
hemisphere, whereas an estimate from the extrapolated magnetic
field shows it to be rather compact along the solar surface
($\lsim 260$~Mm). Therefore, we employ the `parallel' dipole of
\inlinecite{WarmuthMann2005}, which provides a compact domain of
enhanced $V_{\mathrm{A}}$. We adjust the height falloff of the
magnetic field above the active region following
\inlinecite{Gary2001}, but decrease the magnetic field strength to
obtain a realistic $V_\mathrm{fast}$ distribution with model
parameters used.

\begin{figure} 
  \centerline{
                \includegraphics[width=0.5\textwidth,clip=]{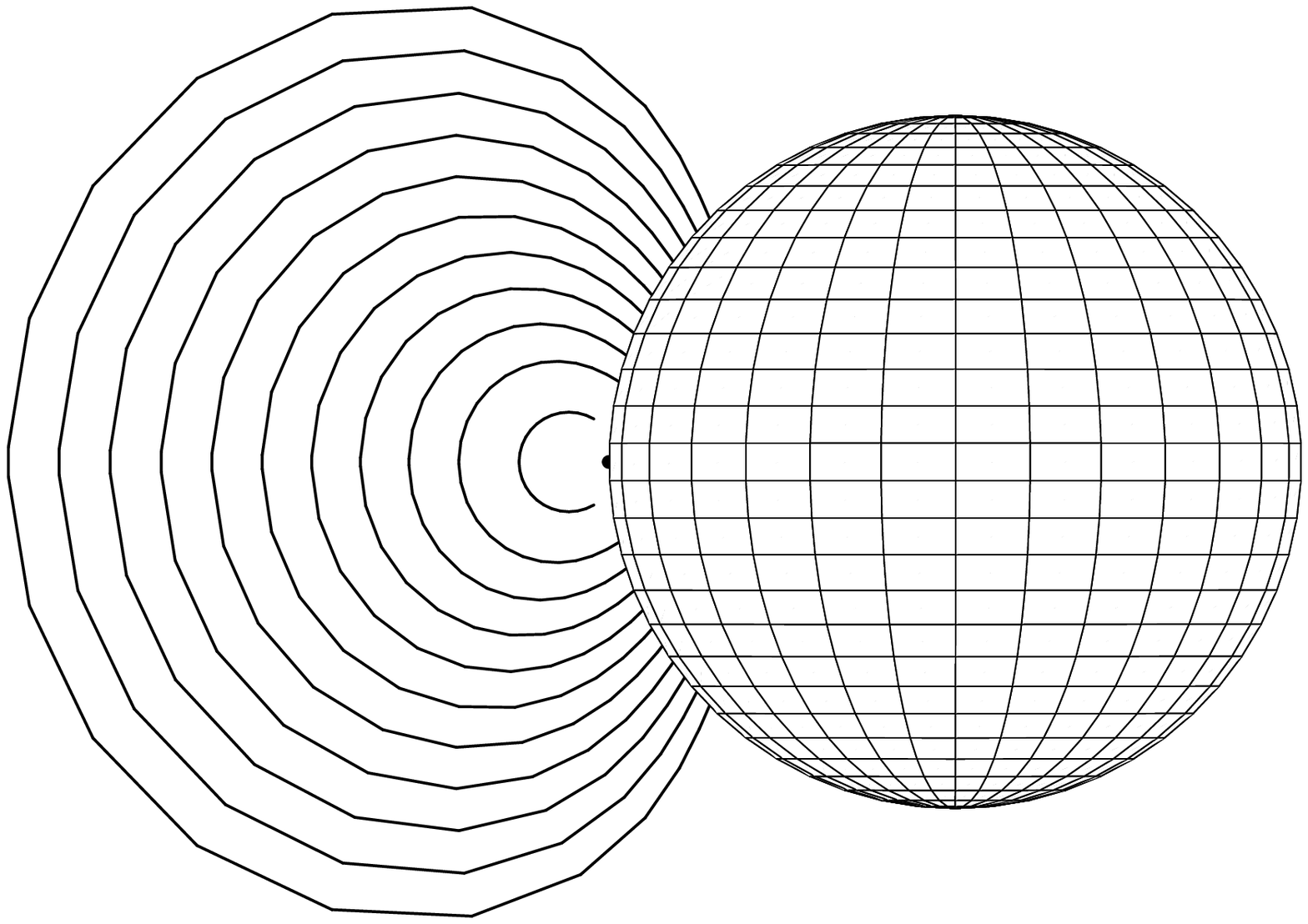}
                \includegraphics[width=0.5\textwidth,clip=]{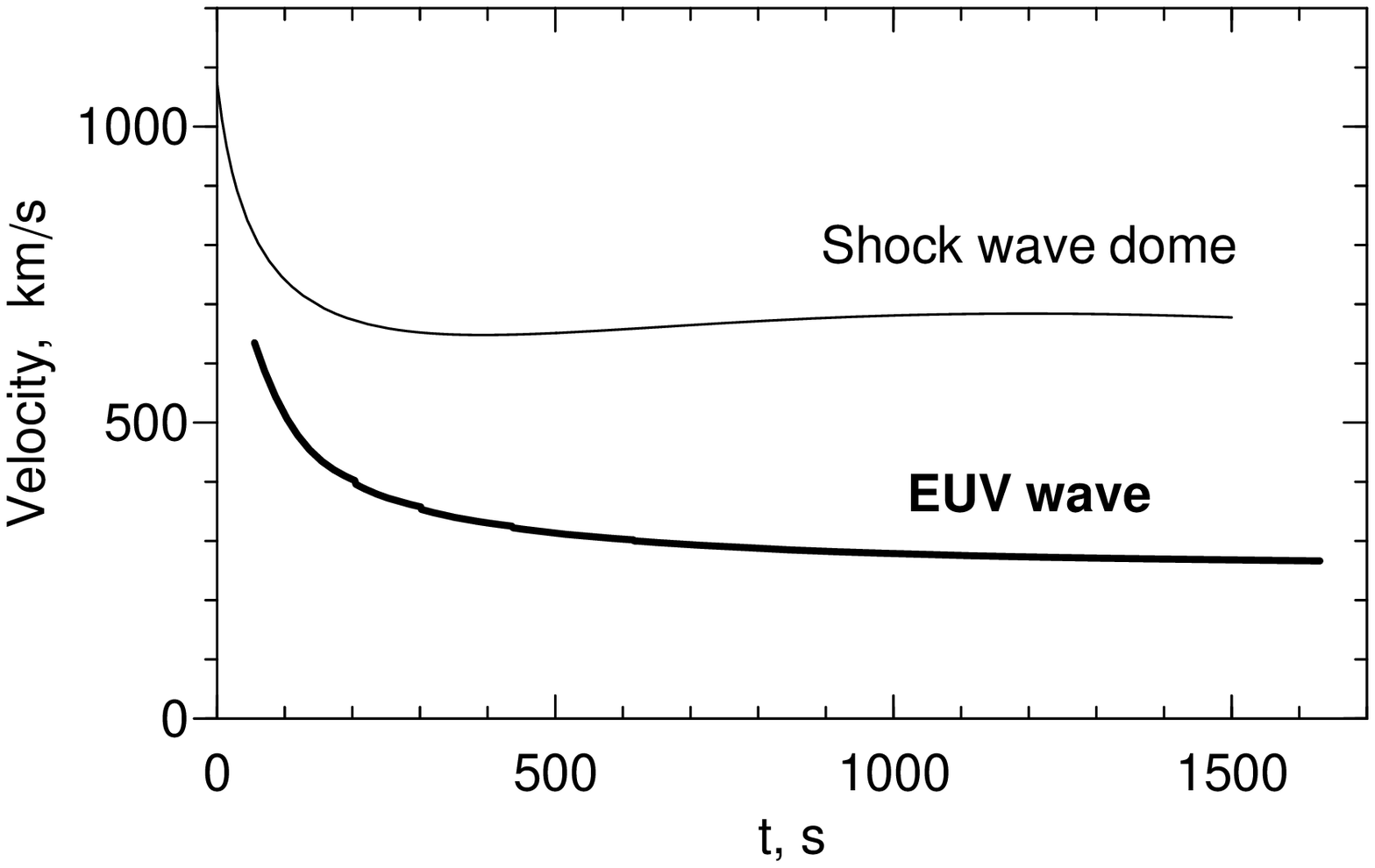}
}
    \vspace{-0.10\textwidth}
          \centerline{\large
      \hspace{0.47 \textwidth}  {a}
      \hspace{0.43 \textwidth} {b}
         \hfill}
   \vspace{0.075\textwidth}

\caption{A freely propagating weak shock wave in the WS model
containing an active region. a)~Shock fronts separated by 2.5-min
intervals. Note the progressive rise of the geometric wave center.
b)~Calculated shock front speeds upwards (thin) and along the
solar surface (thick).
    }
  \label{F-vel_relation}
  \end{figure}

Figure~\ref{F-vel_relation} shows model results. The wave source
is located above the limb in the equatorial plane. The shock front
is oblate in the radial direction presumably due to dominating
upwards increase of $V_{\mathrm{fast}}$: ray trajectories are
refracted into regions of lower $V_{\mathrm{fast}}$. The effect
agrees with the COR1 observations (Figure~\ref{F-cor1}). The speed
of the upwards wave expansion is about twice higher than that of
the on-disk EUV wave (Figure~\ref{F-vel_relation}b). This confirms
our suggestion in Paper~I to overcome the absence of correlation
between the speeds of EUV waves and exciters of type II bursts
stated by \inlinecite{Klassen2000}.

The twice-higher upwards speed of the EUV wave relative to the
on-disk one prompted \inlinecite{Veronig2010} that the upward dome
expansion was driven all the time by the CME. The authors
mentioned that the upward-lateral speed difference could be due to
direction-dependent falloffs of $V_{\mathrm{fast}}$, but preferred
the CME-driven option seemingly favored by the limited lateral
extent of the dimming. However, the latter fact only means that
CME-related opening magnetic fields occurred in a limited region
and did not involve remote regions. The major expansion of all
CMEs is radial, but this fact does not guarantee that all
CME-associated shocks are driven continuously (see Paper~I).

The speed difference in Figure~\ref{F-vel_relation} was obtained
for a \textit{freely propagating wave} and the direction-dependent
$V_{\mathrm{fast}}$ above the active region. The front shapes
match the observations. The results agree with our considerations
and measurements in Paper~I and support the scenario of an
impulsively generated freely propagating weak shock wave (see also
\opencite{PomoellVainioKissmann2008}).

\begin{figure} 
   \centerline{
               \includegraphics[width=0.45\textwidth,clip=,
               bb=109 3 566 308]{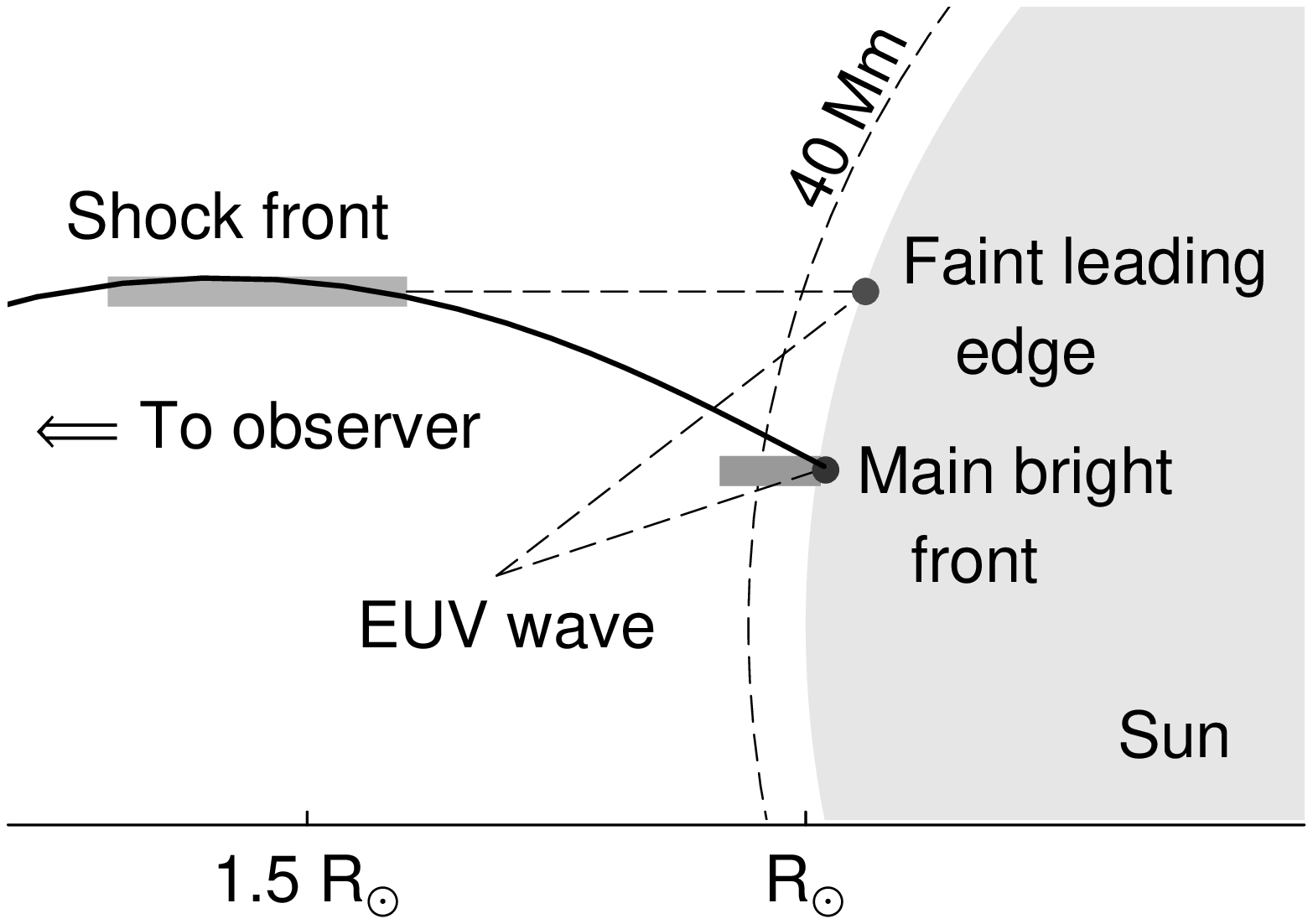}
               \includegraphics[width=0.35\textwidth,clip=]{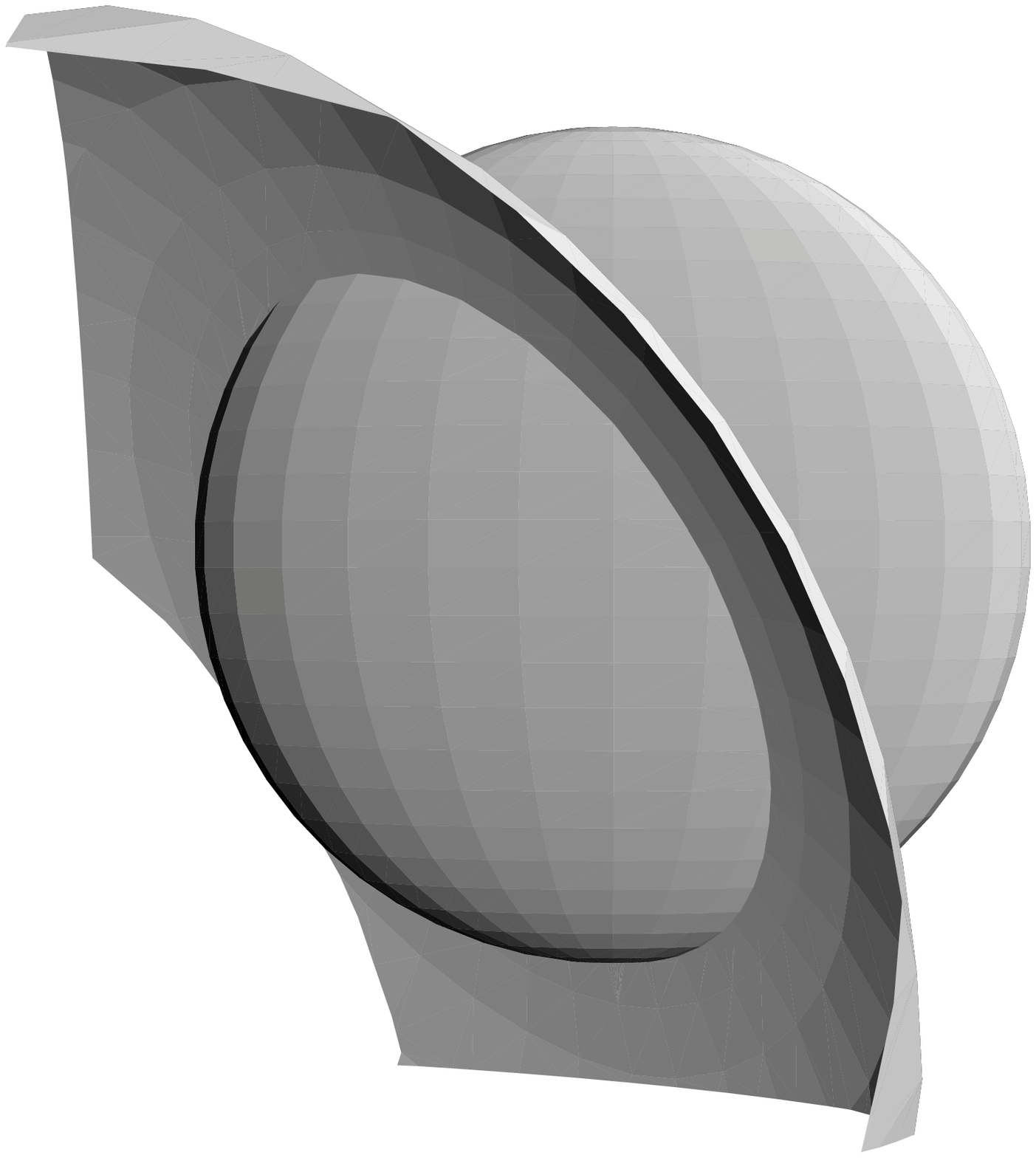}
               \includegraphics[width=0.2\textwidth,clip=]{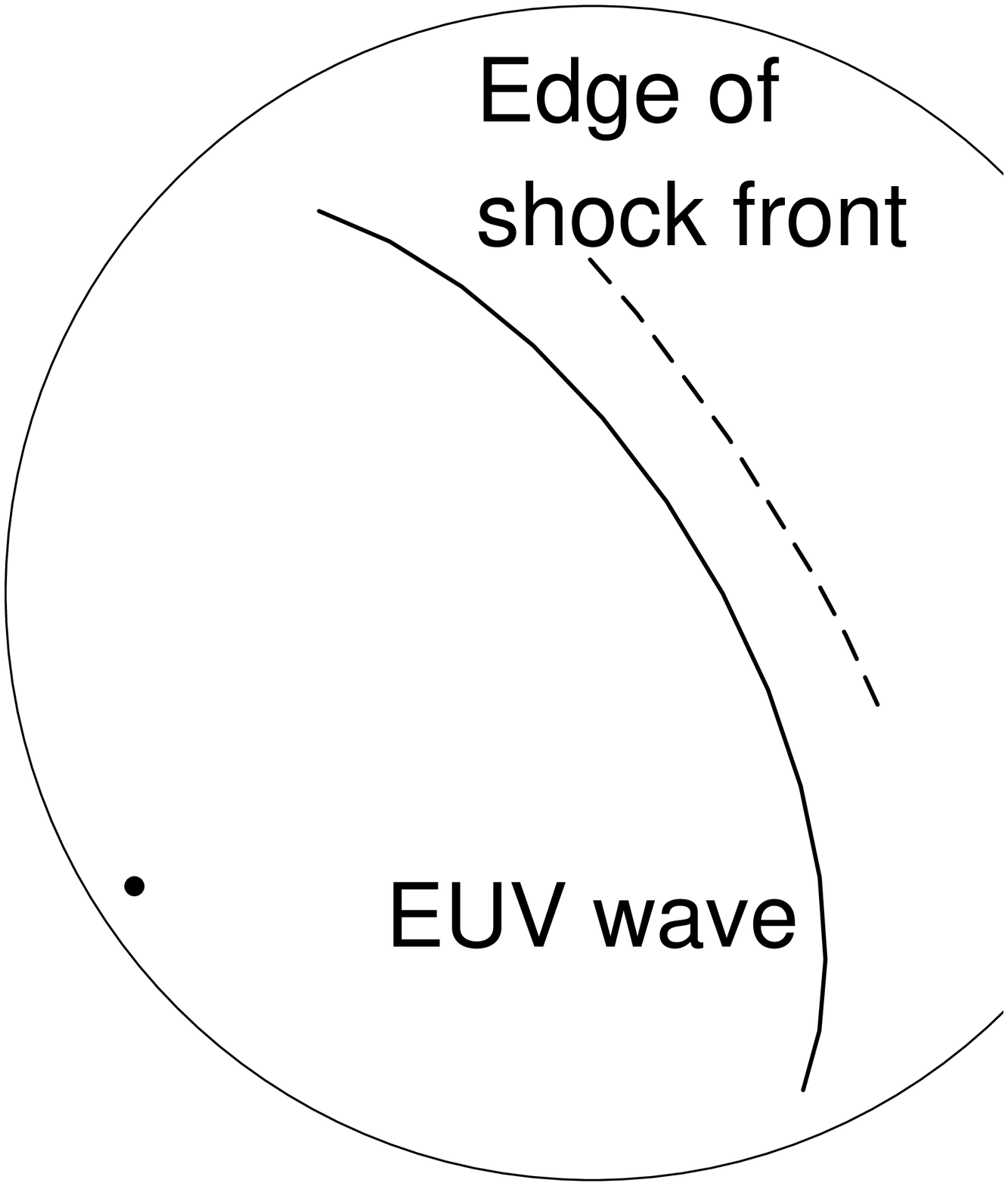}
              }
    \vspace{-0.08\textwidth}
          \centerline{\large
      \hspace{0.03 \textwidth}  {a}
      \hspace{0.42 \textwidth} {b}
      \hspace{0.375 \textwidth} {c}
         \hfill}
   \vspace{0.055\textwidth}
   \caption{Modeling the situation at about 04:30. a)~The relation
between the foremost edge of the EUV wave and its main part. Thick
horizontal bar shows cross section of the wave front presumably
contributing to the foremost edge of the wave detected by Veronig
\textit{et al.} (2010). b)~A portion of the modeled shock front.
c)~The lower edge (solid) of the shock front shown in panel (b)
and a projection of the faint front's foremost edge (broken) on
the solar surface.
        }
   \label{F-wave_front_scheme}
   \end{figure}

The WS modeling explains the disaccord between the EUV wave fronts
identified by us and \inlinecite{Veronig2010}: our red fronts in
Figure~\ref{F-global_euv_wave_fronts} lag behind the blue ones
identified by the authors. The difference is most likely due to a
projection effect combined with a different sensitivity of
measurements as Figure~\ref{F-wave_front_scheme} explains. Plasma
is compressed by the shock front over the whole its surface. The
largest column emission measure of the compression region is near
the solar surface, where the plasma density is higher.
\inlinecite{Veronig2010} probably detected a high-altitude
outermost edge of the convex wave front.
Figure~\ref{F-wave_front_scheme}a demonstrates the calculated 2D
cross section of the wave front with its outermost edge at a
height of $\approx 0.5R_{\odot}$.
Figure~\ref{F-wave_front_scheme}b presents a portion of the
modeled wave front. Figure~\ref{F-wave_front_scheme}c shows the
calculated on-disk projections of the faint leading edge and the
main bright EUV wave front corresponding to about 04:30. The
situation resembles the seemingly disaccord between the results of
\inlinecite{Warmuth2004a} and \inlinecite{WhiteThompson2005}
discussed in Paper~I for a similar reason, \textit{i.e.}, the
convex shape of the wave front and its tilt towards the solar
surface.

\section{Summary and Concluding Remarks}
 \label{S-summary}

Our analysis has confirmed the major conclusion of
\inlinecite{Veronig2010} that both the on-disk EUV wave and the
dome expanding above the limb were due to a coronal shock wave. In
addition to the authors' arguments, we have established that
(1)~the front shape and its changes, (2)~kinematics of both the
on-disk front and the off-limb dome up to $24R_{\odot}$, and even
(3)~the difference between our and the authors' measurements all
corresponded to expected propagation of a shock wave. We have also
found that, in agreement with the shock-wave hypothesis,
kinematics of the global wave front (4)~corresponded to the drift
rate of the type II burst and (5)~was controlled by large-scale
distribution of the fast-mode speed, while its local
inhomogeneities affected the brightness and sharpness of the EUV
wave, \textit{e.g.}, it was brightest in loci of the fast-mode
speed minima.

We do not see any support to the presumption of
\inlinecite{Veronig2010} that the shock wave was driven by the CME
all the time. On the contrary, we consider the shock wave to be
excited by an impulsively erupting magnetic rope structure and
freely propagating afterwards like a decelerating blast wave. This
scenario has been argued and observationally confirmed in Paper~I.
All the conclusions listed in the preceding paragraph are based on
considerations and modeling of freely propagating shock waves. The
free wave propagation is also consistent with the fastest
expansion of its front in EUVI images in the radial direction,
while CME structures apparently lagged behind the wave front.

The shock in this event was most likely weak, at least, near the
solar surface, in agreement with the conclusion of
\inlinecite{Veronig2010}. Model calculations for a weak shock
match observations. Nevertheless, the power-law fit (formally
derived under assumption of a strong self-similar shock wave with
continuously increasing mass) provides reasonable results starting
from the early shock appearance up to latest detectable signatures
of the on-disk EUV wave, and even up to distances $>20R_{\odot}$
from the Sun, although with somewhat variable parameters.

We have additionally revealed another large-scale EUV brightening,
which was quasi-stationary. No manifestations of magnetic field
opening were found outside of the eruption region, while the
propagating on-disk EUV wave was well visible there. The presence
in this event of the two different EUV components predicted by
models offers a promising opportunity to reconcile conflicting
opinions about the nature of ``EUV waves'': the propagating EUV
wave was of a shock-wave nature for sure, and the quasi-stationary
EUV transient was presumably associated with a stretching CME
structure.

We specify the conclusion of \inlinecite{Veronig2010} that the
dome observed in white light was not the CME. Indeed, the leading
part most likely was not a magnetoplasma CME component.
Coronagraph images, their shock-PL fit, and our considerations
indicate that this was a plasma flow successively involved into
the motion by the freely propagating shock front. The plasma flow
was slower than the shock front, whose speed was the phase
velocity of this involvement. Thus, the leading part of the
transient was a plasma flow, \textit{i.e.}, a \textit{coronal mass
ejection} by definition, but it was a \textit{shock-driven plasma
flow}.

\begin{acks}

We thank M. Temmer, A. Warmuth, and P.-F. Chen for fruitful
discussions and S. Kalashnikov for the assistance in data
processing. We thank an anonymous reviewer for useful remarks. We
thank the teams operating all instruments whose data are used here
for their efforts and open data policies: the ESA \& NASA SOHO/EIT
\& LASCO and STEREO/SECCHI telescopes; the NICT HIRAS (Japan), the
IPS Radio and Space Services Learmonth Observatory (Australia),
and the USAF RSTN radio telescopes. We appreciatively use the CME
catalog generated and maintained at the CDAW Data Center by NASA
and the Catholic University of America in cooperation with the
Naval Research Laboratory. SOLIS data used here are produced
cooperatively by NSF/NSO and NASA/LWS. The research was supported
by the Russian Foundation of Basic Research under grant
09-02-00115.

\end{acks}

\end{article}

\end{document}